\documentclass[doublecol]{epl2}

\newcommand{\ff}[1]{{\boldsymbol #1}}
\usepackage{color}

\title{Spin-spin correlations in ferromagnetic nanosystems}
\shorttitle{Title} 

\author{E. Y.\,Vedmedenko\inst{1}\and N.\,Mikuszeit\inst{2}\and T.\,Stapelfeldt\inst{1}\and R.\,Wieser\inst{1} \and M. Potthoff\inst{3}\and A. Lichtenstein\inst{3} \and R. Wiesendanger\inst{1}}
\shortauthor{F. Author \etal}

\institute{
  \inst{1} University of Hamburg, Institute for Applied Physics, Jungiusstr. 11a, 20355 Hamburg, Germany \\
  \inst{2} Instituto Madrile{\~n}o de Estudios Avanzados en Nanociencia, IMDEA-Nanociencia,
Campus  Universidad  Aut{\'o}noma  de  Madrid,  28049  Madrid,  Spain \\
  \inst{3} University of Hamburg, I. Institute for Theoretical Physics, Jungiusstr. 9, 20355 Hamburg, Germany
}


\abstract{
Using exact diagonalization, Monte-Carlo, and mean-field techniques, characteristic temperature scales for ferromagnetic order are discussed for the Ising and the classical anisotropic Heisenberg model on finite lattices in one and two dimensions.
The interplay between nearest-neighbor exchange, anisotropy and the presence of surfaces leads, as a function of temperature, to a complex behavior of the distance-dependent spin-spin correlation function, which is very different from what is commonly expected.
A finite experimental observation time is considered in addition, which is simulated within the Monte-Carlo approach by an incomplete statistical average.
We find strong surface effects for small nanoparticles, which cannot be explained within a simple Landau or mean-field concept and which give rise to characteristic trends of the spin-correlation function in different temperature regimes.
Unambiguous definitions of crossover temperatures for finite systems and an effective method to estimate the critical temperature of corresponding infinite systems are given.
}

\begin{document}

\maketitle

\section{Introduction}

The theory of collective magnetic order is usually concerned with infinite systems \cite{Binder:RepProgrPhys1997,Fisher74,Fisher:PRL72,Wu:PRB76}.
Real magnetic samples, however, have a finite size, and magnetic properties are measured during a finite observation period.
This gives rise to interesting questions as a matter of principle.
For magnetic nanoparticles it also has profound physical consequences.
The recent advances in controlling and measuring magnetic properties of nanoparticles \cite{Krause} as well as applications for magnetic data storage technology \cite{Service} rely on the fact that information, i.e.\ the magnetic state of a finite, small area representing a single bit, is stable over a finite observation time.

The Curie temperature of a ferromagnetic sample is a typical concept that must be reconsidered for nanosized objects.
From the experimental point of view, this is a well-defined quantity which can be measured, e.g.\ as a function of the size $L$ of the nanoparticle \cite{Nepijko,Koetzler:2006}.
The magnetic susceptibility and the specific heat stay finite but show enhancements at $T=T_{\rm C}(L)$, which defines a ``Curie temperature'' up to some residual arbitrariness.

From the theoretical point of view, there is no Curie temperature as there is no phase transition, and actually not even a concept of a thermodynamic phase, in a system with a finite number of degrees of freedom.
Nevertheless, one would like to define $T_{\rm C}(L)$ roughly to be the temperature where the ferromagnetic alignment of the spins within the particle becomes stable against thermal fluctuations.
In a simple mean-field picture, one finds $T_{\rm C}(L) < T_{\rm C}(\infty)$ where $T_{\rm C}(\infty)$ is the precisely defined Curie temperature of the corresponding (infinite) bulk system.

The transition, or smooth crossover, from a paramagnetic (PM) to an ordered state in the nanosystem must be described within models of interacting \emph{microscopic} (atomic) spins \cite{Landau:PRB1976,Dimitrov:PRB96}.
Below $T_{\rm C}(L)$ the different microspins are tightly bound together and form a huge \emph{macrospin}\cite{Koetzler:2006,Dimitrov:PRB96}.
For a finite-size and spin-isotropic system, the direction of the macrospin fluctuates strongly, i.e.\ the magnetic state of the system is not stable temporally.
Anisotropies give rise to superparamagnetic (SPM) behavior for temperatures above the so-called blocking temperature $T_{\rm b}=T_{\rm b}(L)$, i.e.\ for $T_{\rm b}(L)<T<T_{\rm C}(L)$.
Both quantities are size-dependent.
It is rather the blocking temperature $T_{\rm b}(L)$ than $T_{\rm C}(L)$ that is relevant for storage technology since it characterizes the crossover from the stable ferromagnetic (FM) state at low $T$ to the SPM state where the system ``switches'' between different energy minima determined by magnetic anisotropies.

The topic is complicated by the fact that $T_{\rm b}(L)$ cannot be considered as a pure property of the system.
It must be seen as a relative value, which depends on the observation time $\tau$.
For $\tau\rightarrow\infty$, there is no blocking of the magnetization as eventually the anisotropy energy barrier is overcome by thermal fluctuations or even due to quantum tunneling, and hence $T_{\rm b}(L)\rightarrow 0$.
For $\tau\rightarrow 0$ (referring to e.g.\ laser-probe methods), $T_{\rm b}(L) \rightarrow T_{\rm C}(L)$ while for intermediate $\tau$ (like in spin-polarized scanning tunneling microscopy \cite{Wiesendanger2009}) $0<T_{\rm b}(L) < T_{\rm C}(L)$.

For the infinite system, a magnetic phase transition is characterized by a divergence of the correlation length $\xi$, which characterizes the spatial decay of the spin-spin correlation function.
For a nanosized system, on the other hand, it is by no means clear how the crossover at the Curie temperature $T_{\rm C}(L)$ manifests itself in the correlation function.
The purpose of the present paper is to provide a systematic study of the spin-spin correlation function for isotropic and anisotropic classical spin models in different dimensions with a finite and, in a thermodynamic meaning, {\em small} number of microspins.
To this end, we compare results obtained by Landau mean-field theory, exact diagonalization, and Monte-Carlo data.
As our main result we find that three different temperature scales, the blocking temperature $T_{\rm b}(L)$ and the Curie temperature $T_{\rm C}(L)$ of the finite spin system as well as the Curie temperature of the infinite bulk $T_{\rm C}(\infty)$, can be read off from a suitably defined spin correlation function, which is accessible to scattering experiments.
We propose a simple three-parameter fit formula for the correlation function, which turns out to be very effective in describing the numerical data for the entire temperature range and may serve to give a definition for the Curie temperature that is consistent with the usual estimates of $T_{\rm C}(L)$ based on the magnetic susceptibility or the specific heat.
Finally, the blocking temperature, relative to the observation time, can easily be accessed by interpreting the Monte-Carlo sweeps as time steps.
Our results demonstrate that studies based on microscopic spin models in the superparamagnetic regime must go beyond the mean-field level.

\section{Spin correlation function}

The correlation function between two spins $\ff S_i$ and $\ff S_j$ at sites $\ff r_i$ and $\ff r_j$ is given by
\begin{eqnarray}\label{eq:gamma0}
  G(\ff r) &=& \langle \ff S_i \ff S_j \rangle \: ,
\end{eqnarray}
where $\langle \cdots \rangle$ is the canonical thermal average at temperature $T$.
For a translationally invariant bulk system, the correlation function is homogeneous and depends on the translation vector $\ff r = \ff r_i - \ff r_j$ only, while in case of a finite system it depends on $\ff r$ and on the reference site in addition.
For the following discussion we define an averaged correlation function which is independent of the direction and depends on the distance $r=|\ff r|$ only:
\begin{equation}\label{eq:gamma1}
  G(r) = \frac{1}{n(r)}\sum\limits_{\scriptstyle   i < j \atop
  \scriptstyle |{{\ff r}_i} - {{\ff r}_j}|=r }\big \langle \ff S_i \ff S_j \big\rangle  \: .
\end{equation}
Here, the sum in the first term runs over all $n(r)$ pairs separated by the distance $r$.
$G(r)$ directly refers to X-ray- or neutron-scattering experiments.
Furthermore, we define a ``connected'' correlation function
\begin{equation}\label{eq:gamma2}
  \widetilde G(r) = G(r) - M^2 \; ,
\end{equation}
where $M = | \sum_i \langle \ff S_i \rangle | / L$ and $L$ being the number of sites.
Apart from a constant factor, $\ff S_i$ is the local magnetic moment at site $\ff r_i$, and thus $M$ is the magnetization of the nanosystem.
For temperatures above the blocking temperature, $M$ averages to zero.
But even below $T_b(L)$ the magnetization vanishes, $M=0$, in an exact calculation. The reason for the vanishing magnetization is the infinitely long "observation time" or time averaging in exact calculations.
If, on the other hand, the average $\langle \cdots \rangle$ is interpreted as a time average and if the time constant defining the average (life-time of a state) is large as compared to the observation time $\tau$, we have a finite $M$ for temperatures below the blocking temperature corresponding to $\tau$. Hence, the function $\widetilde G(r)$ will strongly differ from $G(r)$ or even vanish.

\section{Ising model}

\begin{figure} \centering
\includegraphics*[width=0.67 \columnwidth]{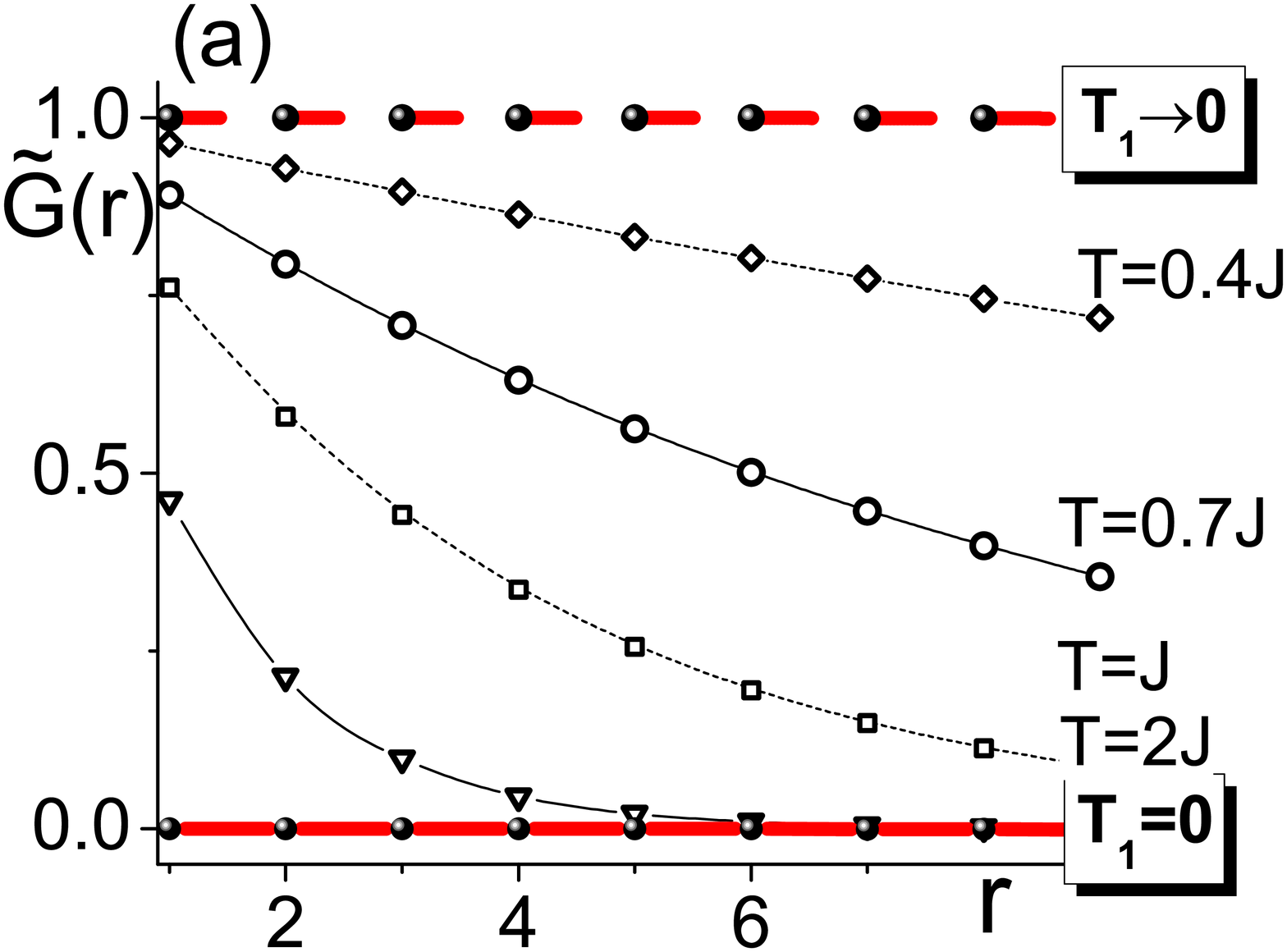}

\includegraphics*[width=0.82 \columnwidth]{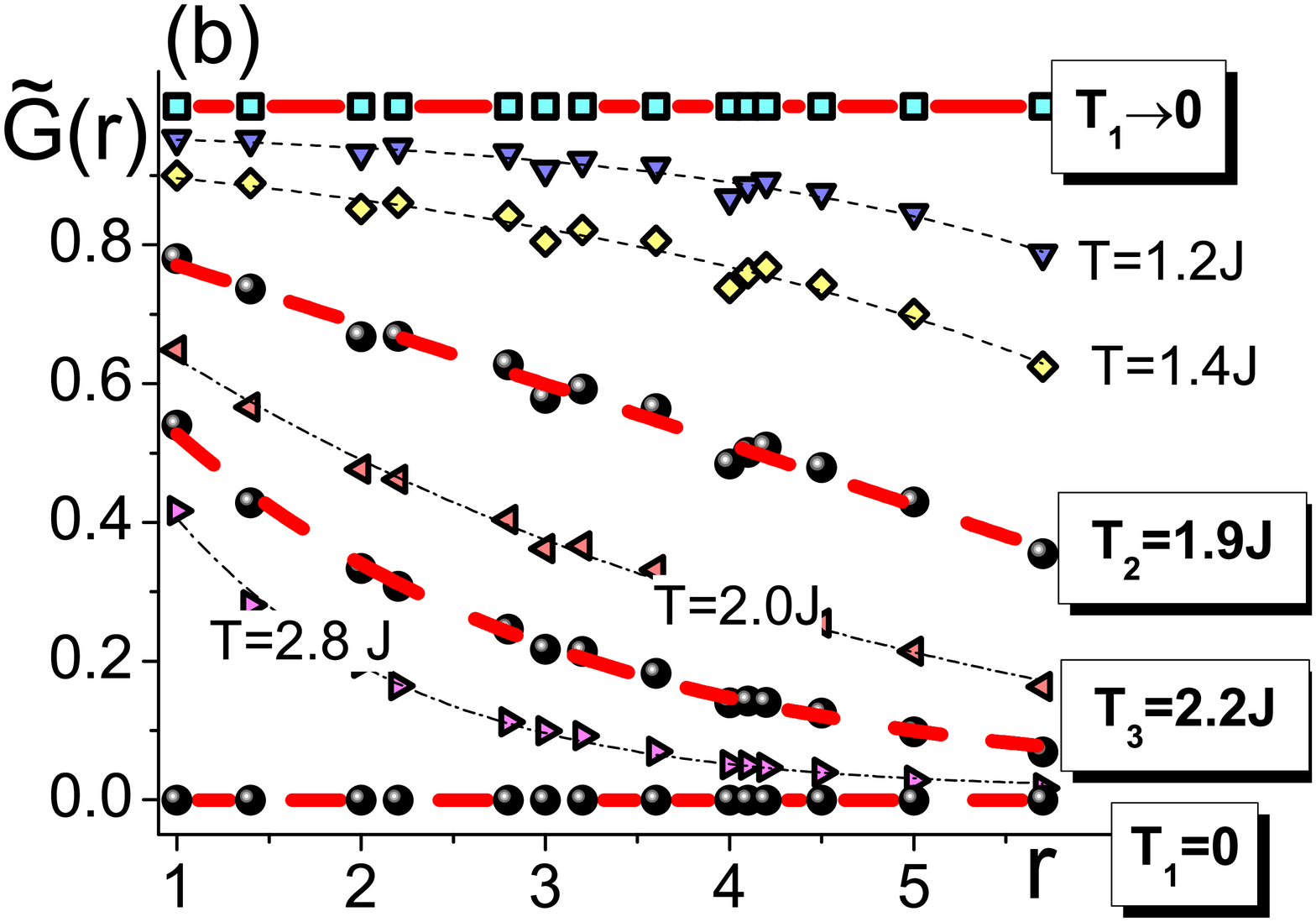}
\caption{Spin correlation function $\widetilde G(r)$ (see Eq.\ (\ref{eq:gamma2}), symbols) calculated for an open Ising chain consisting of $L=10$ sites (a) and for an open $5\times 5$ Ising square lattice (b). The data are fitted with Eq.\ (\ref{eq:PairCorr}). The temperatures $T_1$ (a) and $T_1$, $T_2$, $T_3$ (b) are highlighted by thick red lines.
}
\label{fig:Gamma}
\end{figure}

We start the discussion by considering the ferromagnetic ($J>0$) Ising model:
\begin{equation}\label{eq:ising}
  H = - J \sum_{\langle i,j \rangle} S_{iz} S_{jz}  \; ,
\end{equation}
on a one-dimensional chain of length $L$ with open boundary conditions.
The sum runs over all pairs of nearest-neighbor sites.
The average correlation function $\widetilde G(r)$ is easily calculated analytically.
Results for $L=10$ and different temperatures are displayed in Fig.\ (\ref{fig:Gamma}a).
We note that there is a simple exponential decay of the correlations, $\widetilde{G}(r) \propto \exp(-r/\xi)$, on a length scale $\xi$ which at low temperatures exceeds the system size.
Although in the low-temperature regime the microspins are perfectly aligned ferromagnetically and although they become uncorrelated on length scales much smaller than the system size in the high-temperature limit, there is no meaningful ``Curie point'' that could be extracted from $\widetilde G(r)$.
Qualitatively, not much happens as a function of $T$.
The absence of a Curie point, $T_{\rm C}(L)=0$, is of course not unexpected.
It corresponds to a featureless magnetic susceptibility $\chi(T)$ and to the fact that $T_{\rm C}(\infty)=0$ for the infinite Ising chain.
Only at $T=0$, the system ``freezes'' in one of the two ferromagnetic ground states, i.e.\ $M=1$, and therefore, due to our definition (\ref{eq:gamma2}), the correlation function discontinuously jumps to $\widetilde G(r) = 0$.
This might be expressed as a vanishing blocking temperature, $T_{\rm b}(L)=0$, reflecting the fact that an exact calculation corresponds to an infinite observation time $\tau$.

\begin{figure} \centering
\includegraphics*[width=1.0  \columnwidth]{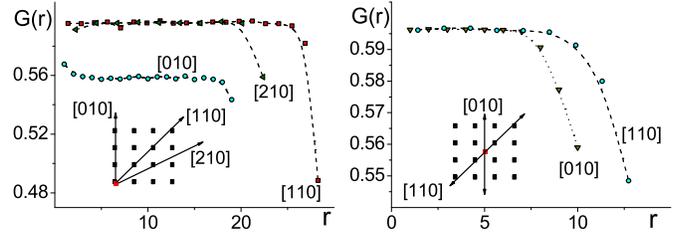}
\caption{Spin correlation function $G(\ff r)$ (see Eq.\ (\ref{eq:gamma0}), symbols) calculated for a $20\times 20$ Ising square lattice with open boundaries starting from an edge (left) and from the central site (right) as obtained by Monte-Carlo simulations at $T=0.5J \ll T_{\rm C}(\infty)$.
The lines are guides to the eyes only.
}
\label{fig:Gamma1}
\end{figure}

For a finite two-dimensional Ising array with $L = 5 \times 5$ sites, the situation changes completely.
Calculations for $L=25$ are easily done by numerically exact diagonalization.
Results obtained for the average correlation function $\widetilde G(r)$ [Eq.\ (\ref{eq:gamma2})] are shown in Fig.\ \ref{fig:Gamma}b.
The result is surprising:
We find two or, including $T=0$ (see below), three different crossover temperatures.

For high temperatures, the correlations decay exponentially, see $T=2.8 J$, for example.
Below a temperature $T_3$, however, the trend can no longer be fitted by an exponential of the form $\exp(-r/\xi)$.
We find $T_3 \approx 2.2 J$.
This is close to the bulk Curie temperature of the two-dimensional Ising model $T_3 \approx T_{\rm C}(\infty) = 2/\ln(1+\sqrt{2}) J \approx 2.27 J$.
Upon lowering $T$ we then find another temperature $T_2 \approx 1.9 J$ which is characterized by a change of the curvature of $\widetilde G(r)$.
Below $T_2$, the trend of $\widetilde G(r)$ is no longer convex but concave until $r$ hits the system boundary.
As the exact-diagonalization data correspond to an infinite observation time, the third temperature scale $T_1$ is trivially given by the vanishing blocking temperature $T_1=T_{\rm b}(L)=0$.

\section{Model correlation function}

To extract the different temperature scales $T_3, T_2, T_1$, we propose to fit the correlation function to the following expression with three temperature-dependent parameters:
\begin{equation}\label{eq:PairCorr}
  \widetilde{G}(r) = \widetilde{G}(r,T) \approx B(T) e^{-r/\varepsilon(T)} + y(T) \: .
\end{equation}
In this way $T_3$ is defined by the temperature where $y(T)$ becomes finite, i.e. where a deviation from a purely exponential decay of $\widetilde{G}(r)$ is found.
For the infinite system, $L\to \infty$, this happens right at $T_3=T_{\rm C}(\infty)$ where a power-law decay is expected \cite{Fisher74,Campanino:EPL2003}.
Thereby, the simple fit formula, Eq.\ (\ref{eq:PairCorr}), will provide a rough estimate for $T_{\rm C}(\infty)$ based on a single calculation of $\widetilde{G}(r)$ for a finite system.
Our Monte-Carlo calculations show that this estimate reliably gives $T_{\rm C}(\infty)$ within an error of the order of 1\%.
For example, using the fit for Monte-Carlo data obtained for the $5 \times 5$ Ising array, we find $T_3 = 2.20 \pm 0.02$ while for the $8\times 8$ lattice $T_3 = 2.26 \pm 0.02$.
This represents a cheap but rough way to get $T_{\rm C}(\infty)$ from a slow annealing of a {\em single} finite system.
For temperatures below the blocking temperature, we have $\widetilde{G}(r) = 0$ ($G(r)=1$).
Hence, $T_1=T_{\rm b}(L)$ is indicated by $B(T) = 0$ and $y(T) = 1$ when fitting the data using Eq. (\ref{eq:PairCorr}).
The main purpose of Eq. (\ref{eq:PairCorr}), however, is to get an estimate for $T_2$ where $\widetilde{G}(r)$ is a linear function to a good approximation.
A linear $\widetilde{G}(r)$ requires $\varepsilon(T) \to \infty$ for $T\to T_2$.
To get a finite slope, we also need $B(T) \to \infty$ for $T\to T_2$, and finally $y(T) \to \infty$ since $|G(r)|\le 1$.

We have used Eq.\ (\ref{eq:PairCorr}) to fit the unknowns $\varepsilon(T)$, $B(T)$ and $y(T)$ to numerically exact data for one-, two- and three-dimensional Ising systems of different size $L$ as well as for finite isotropic and anisotropic Heisenberg systems.
It turns out that the quality of the fit is exceptionally good in the entire temperature range, see the lines in Fig.\ \ref{fig:Gamma}a and b, for example.
For temperatures $T_{\rm b}(L)<T<T_2$ we find the concave trend with negative values for $B(T)$ and $\varepsilon(T)$ while $y(T)>0$.
For $T_2<T<T_{\rm C}(\infty)$, both $B(T)$ and $\varepsilon(T)$ are positive while $y(T)<0$, and the trend of $G(r)$ is convex.
Finally for $T>T_3$ both $B(T)$ and $\varepsilon(T)$ remain positive while $y(T)$ vanishes
leaving an exponential decay of $\widetilde{G}(r)$.

\section{Curie temperature}

We identify $T_2$ with the ``Curie temperature'' of the system, $T_2=T_{\rm C} (L)$.
This provides a meaningful definition of $T_{\rm C} (L)$ for a finite system that is based on the spin correlation function.
It is motivated by the physical idea that at $T_{\rm C}(L)$ the correlation length exceeds the system size but additionally takes into account that the nanosystem is bounded by surfaces.
For an infinite system the spin correlation function is always convex, i.e. its slope is negative but increasing as a function of increasing distance $r$.
The unusual concave trend of the correlation function at lower temperatures must therefore be a direct consequence of the presence of surfaces.
This is demonstrated with Fig.\ \ref{fig:Gamma1} for a larger system consisting of $20 \times 20$ Ising spins on a square array.
Due to missing nearest neighbors at the nanoparticle surface, fluctuations of the local spins are stronger and result in a reduced average surface magnetization.
This also implies a strongly decreasing correlation function $G(\ff r)$ (see Eq.\ (\ref{eq:gamma0})) close to the surface and along any direction.
For the averaged correlation function $\widetilde{G}(r)$ (see Eq.\ (\ref{eq:gamma1})), this surface effect competes with the convex bulk trend of $\widetilde{G}(r)$.
The surface effect dominates for $T>T_{\rm C}(L)$ and drives the nanosystem to a ``paramagnetic'' state while for $T<T_{\rm C}(L)$ the bulk of the nanosystem causes an ordered superparamagnetic state, and the surface manifests itself in stronger fluctuations of the spins and a concave trend of $\widetilde{G}(r)$ only.
Note that the fit with the model correlation function, Eq.\ (\ref{eq:PairCorr}), allows to characterize $T_{\rm C}(L)$ by a divergence of the parameter $\varepsilon(T)$, which therefore might be called a ``virtual correlation length''.
On the other hand, the Curie temperature of the infinite system $T_{\rm C}(\infty)$ is given by a divergence of
$r_{\rm v} \equiv - \varepsilon(T) \ln[-y(T)/B(T)]$ since $y(T)$ becomes finite at $T_{\rm C}(\infty)$ in the fit.
$r_{\rm v}$ is the distance at which $\widetilde{G}(r)$ vanishes, $\widetilde{G}(r_{\rm v})=0$.
The distance is ``virtual'' because it is always larger than the system size (see Fig.\ \ref{fig:GammaMC}a).

\section{Blocking temperature}

\begin{figure} \centering 
\includegraphics*[width=0.47 \columnwidth]{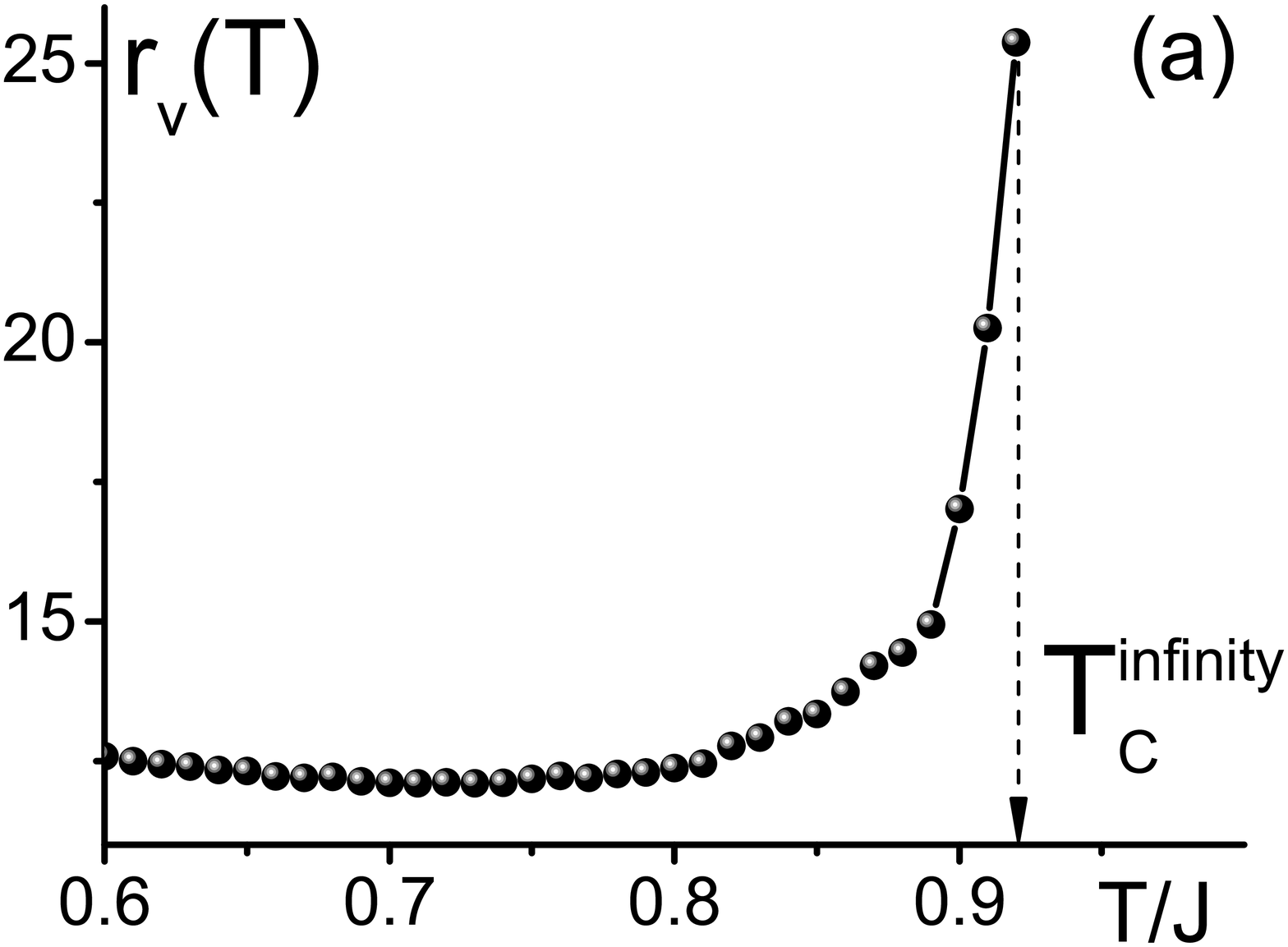}
\includegraphics*[width=0.47 \columnwidth]{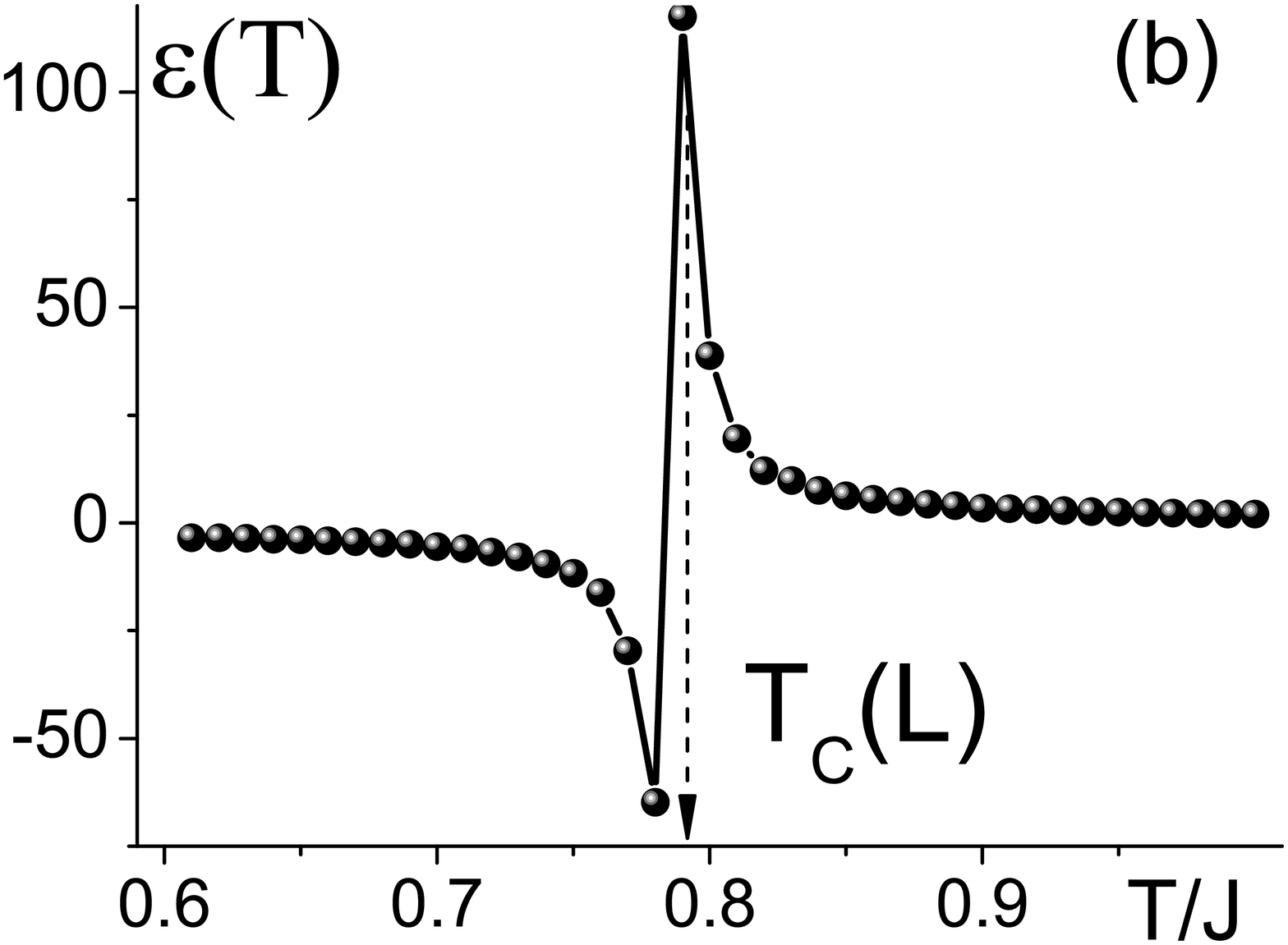}
\\[8mm]
\includegraphics*[width=1.0 \columnwidth]{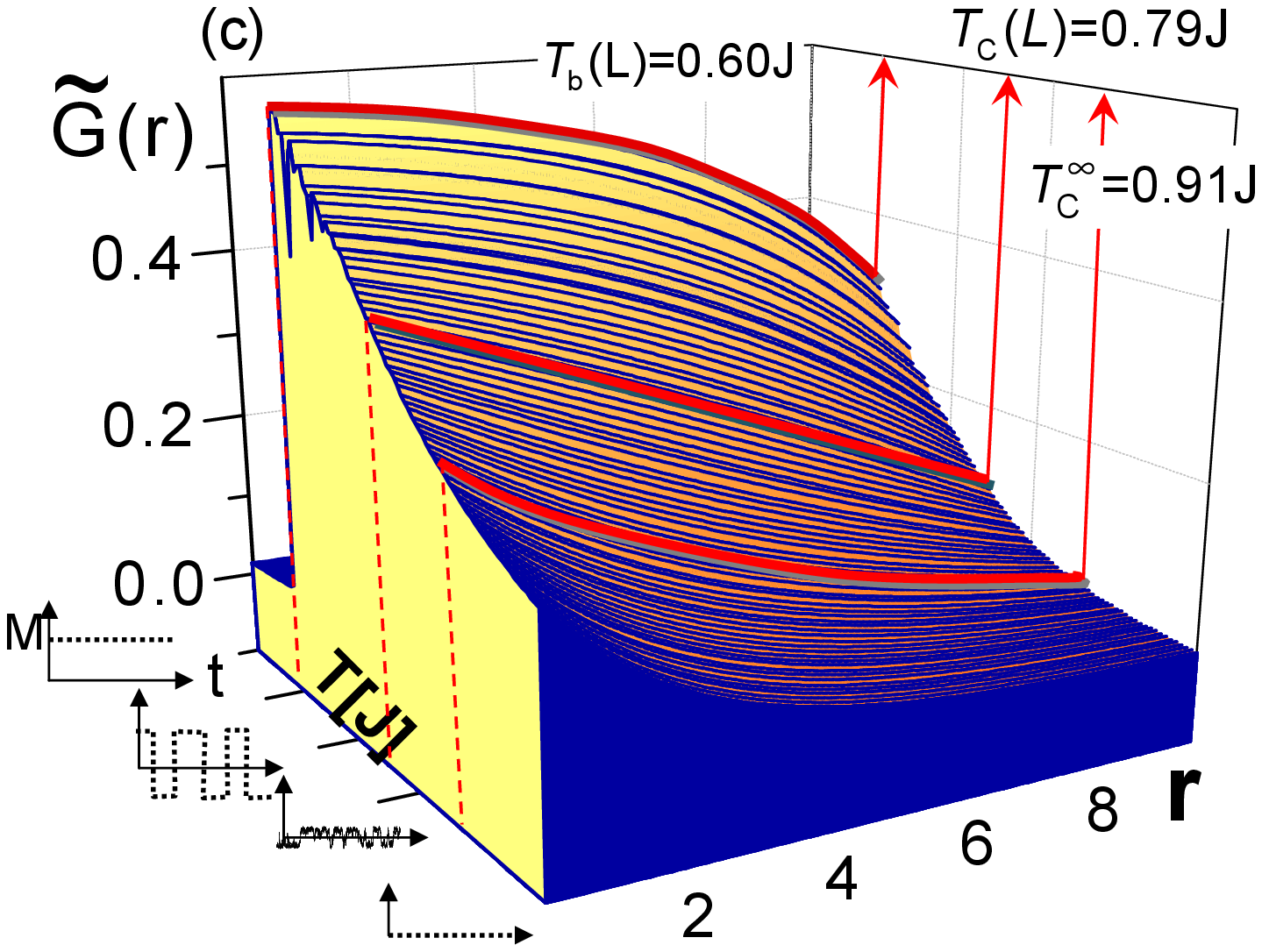}
\caption{Model $\widetilde{G}(r)$ [see Eq.\ (\ref{eq:PairCorr})] as obtained by fitting the parameters to results of Monto-Carlo simulations ($10^6$ sweeps per temperature) for an anisotropic Heisenberg model ($K=0.6J$) on an $8\times 8$ square lattice with open boundary conditions.
(a) Temperature dependence of the virtual distance $r_v$ (see text).
(b) Virtual correlation length $\varepsilon(T)$.
(c) $\widetilde G(r)$ as a function of $r$ and $T$.
The temperatures $\infty > T_{\rm C}(\infty) > T_{\rm C}(L) > T_{\rm b}(L) > 0$ are highlighted.
In each of the four corresponding temperature ranges, the typical Monte-Carlo time dependence of $M=M(t) \propto \langle \sum_i S_i^z\rangle$ is shown in insets (see text for discussion).
}
\label{fig:GammaMC}
\end{figure}

The question of a finite blocking temperature $T_{\rm b}$ can be addressed by Monte-Carlo simulations when interpreting Monte-Carlo sweeps as time steps \cite{Vedmedenko}.
A finite number of Monte-Carlo sweeps corresponds to an incomplete statistical average and thus to a finite observation time $\tau$.
Calculations have been performed for the classical Heisenberg model with nearest-neighbor exchange $J>0$ and uniaxial magnetic anisotropy $K>0$ on finite two-dimensional square lattices:
\begin{equation}
  \label{eq:H}
  H = -J \sum_{\langle ij \rangle} {\ff S}_i \ff{S}_j - K \sum\limits_{i} \left(S_{i}^z \right)^2 \: .
\end{equation}
Each Monte-Carlo run consists of up to 200 temperatures with up to $10^7$ sweeps per temperature.

An example for $K=0.6J$ is given in Fig.\ \ref{fig:GammaMC}c.
The fit of the Monte-Carlo results by Eq.\ (\ref{eq:PairCorr}) is accurate {\em for all temperatures and all values of $K$} such that the three different temperature scales, $T_{\rm b}(L), T_{\rm C}(L)$, and $T_{\rm C}(\infty)$ can be extracted easily.
The finite blocking temperature manifests itself in the jump of $\widetilde{G}(r)$ as a function of $T$ which is due to the jump of $M$ at $T \approx 0.6 J$.
The order of magnitude for $T_{\rm b}$ seems to be given by $K$.
However, $T_{\rm b}(L)$ decreases with increasing observation time $\tau$, i.e.\ with an increasing number of Monte-Carlo sweeps per temperature.
At fixed $L$, we find $T_{\rm b}(L)\rightarrow 0$ logarithmically if $\tau \to \infty$.

Fig.\ \ref{fig:GammaMC}c nicely demonstrates that an astonishingly complex behavior of the spin correlation function is found for finite anisotropic nanosystems.
The qualitatively different physics within the different temperature ranges, i.e.\
$0 < T_{\rm b}(L) < T_{\rm C}(L) < T_{\rm C}(\infty)<\infty$, also shows up in the qualitatively different behavior of the order parameter $M$ as a function of (Monte-Carlo) time during the simulation, see insets in Fig.\ \ref{fig:GammaMC}c:
Below $T_{\rm b}(L)$, on the scale set by the observation time, the magnetization freezes in one of the values corresponding to the degenerate energy minima of the anisotropic model; for $T_{\rm b}(L)<T<T_{\rm C}(L)$ the magnetization switches between these values with a switching time which is much smaller than the ``magnetic lifetime'' of a state; for $T_{\rm C}(L)<T<T_{\rm C}(\infty)$ the system still switches but the lifetimes are comparable to the switching times; and finally above $T_{\rm C}(\infty)$ correlations decay exponentially and $M=0$.

\section{Dependence on the system size}

%
%
%

Note that this physics is characteristic of a {\em finite} system:
For constant $\tau$ but increasing system size $L \to \infty$ all three temperatures merge, and we are left with a single critical temperature only, the Curie temperature of the infinite system $T_{\rm C}(\infty)$.
With increasing $L$, but keeping the temperature fixed, the curvature of $\widetilde{G}(r)$ increases, i.e.\ it becomes less convex, changes from convex to concave, or becomes more concave.
This is due to the less and less important effect of the nanoparticle's surfaces.
At the same time $\widetilde{G}(r)$ and its slope increase.
This implies that $T_{\rm C}(L)$ is an increasing function of the system size.
The same holds for the blocking temperature since with increasing $L$ the energy of the anisotropy barrier increases and higher temperature is needed to induce a thermal switching of the magnetization.

\begin{figure} \centering
\includegraphics*[width=0.95\columnwidth]{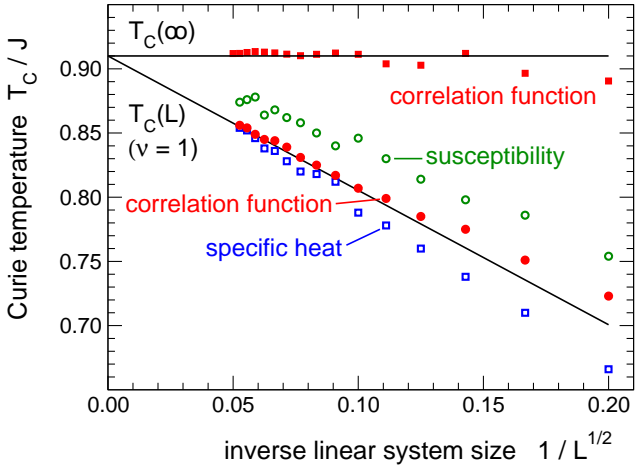}
\caption{
Curie temperature, as obtained from the maximum of the specific heat (open squares), from the maximum of the susceptibility (open circles) and from the correlation-function fit (filled circles), as functions of the (inverse) linear system size, and bulk Curie temperature (filled squares), as obtained from the fit of the correlation-function at the respective system size.
Solid line: $(T_{\rm C}(\infty) - T_{\rm C}(L) ) / T_{\rm C}(\infty) = (L/L_0)^{-1/2}$ corresponding to the exponent $\nu = 1$ and with $L_0^{1/2}=1.15$.
Calculations for the $D=2$ anisotropic Heisenberg model with $K=0.6 J$.
}
\label{fig:tcofl}
\end{figure}

\begin{figure} \centering
\includegraphics*[width=0.96\columnwidth]{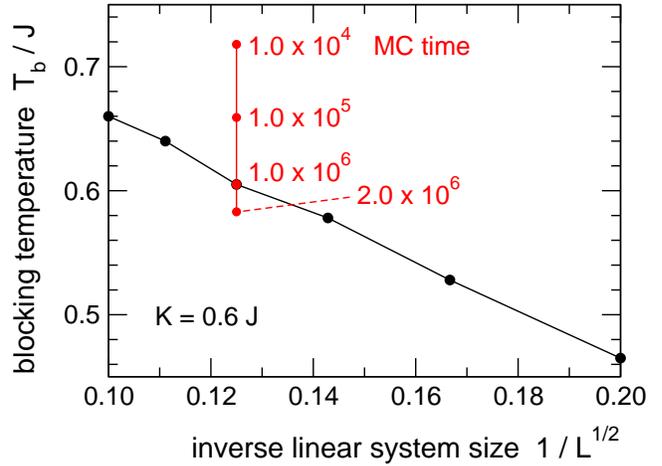}
\caption{
Blocking temperature, as obtained from the correlation-function fit, as a function of the (inverse) linear system size for the $D=2$ anisotropic Heisenberg model with $K=0.6 J$ and using $1.0 \cdot 10^6$ Monte-Carlo sweeps.
Lines connect the data points.
For the $L=8\times 8$ system, the dependence of $T_{\rm b}$ on the Monte-Carlo time is given by the red circles as indicated.
}
\label{fig:tblocking}
\end{figure}

The system-size dependence of the Curie temperature, as obtained from the fit of the correlation function, is displayed in Fig.(\ref{fig:tcofl}). Comparing $T_{\rm C}(L)=T_2$ with the Curie temperature defined by the maximum of the magnetic susceptibility and by the maximum of the specific heat, we find that the asymptotic behavior is approached significantly faster when using the correlation-function fit.
The latter also provides a reliable estimate for the bulk Curie temperature as is also shown in  Fig.\ (\ref{fig:tcofl}).
In addition, Fig.\ (\ref{fig:tblocking}) shows the dependence of the blocking temperature on the system size.
For fixed system size and with increasing Monte-Carlo time, i.e.\ with increasing number of sweeps, the blocking temperature decreases logarithmically (see data for $L=8\times 8$ in Fig.\ (\ref{fig:tblocking})).

Our analysis shows that $T_{\rm C}(L)$ satisfies the finite-size scaling law $(T_{\rm C}(\infty)-T_{\rm C}(L))/T_{\rm C}(\infty)=(L/L_0)^{-1/D\lambda_s}$ \cite{barber,Iglesias}.
$\sqrt{L_0}$ corresponds to a microscopic length scale, and its order of magnitude is one.
The shift exponent $\lambda_s$ is related to the exponent of the correlation function via $\lambda_s = 1/ \nu$.
In case of the anisotropic Heisenberg model, our data for $T_{\rm C}(L)$ for system sizes up to $L=19\times 19=361$ are consistent with $\nu=1.0$ (and $\sqrt{L_0}=1.15$).
This is different from the classical exponent ($\nu = 0.5$) and agrees with the exponent for the $D=2$ Ising model ($\nu=1$).

Note, that our results for the size dependence of the blocking temperature are also consistent with the same scaling law that describes $T_C(L)$, see Fig.\ (\ref{fig:tblocking}).
The exponent $\lambda_s$, however, is different and slightly larger than $1/\nu$ for $\nu=1$.
Actually, it is by no means clear that the blocking temperature should satisfy a scaling law since for the {\em infinite} system it has no meaning independent from the Curie temperature.
This point is beyond the scope of the present study but deserves further investigations.

\section{Static mean-field theory}

The characteristic trends of the spin-correlation function in the different temperature regimes are strongly determined by the presence of surfaces, see Fig.\ \ref{fig:Gamma1}.
It is tempting to simply explain the concave trend of the correlation function at temperatures below $T_{\rm C}(L)$ by the reduced coordination numbers at the nanoparticle surfaces.
We have checked this by performing calculations using periodic boundary conditions.
In fact, a convex curvature of the correlation function is found in one and in two dimensions and for all temperatures in agreement with previous work \cite{Landau:PRB1976}.
Since the surface-to-volume ratio is smaller in one as compared to two dimensions, the simple coordination-number argument is also consistent with the absence of a Curie temperature in one dimension.
We have checked our findings by performing corresponding calculations for three-dimensional finite lattices:
In fact, the low-temperature concave trend is found to be even more pronounced here.

Simple coordination-number arguments are included in the Landau theory of magnetic systems bounded by surfaces
\cite{Mills1971,Binder1972,Kumar1974,Lubensky} where the usual Landau free-energy functional in considered but with an additional surface free-energy term.
The resulting Landau mean-field theory is essentially equivalent to static mean-field theory for a discrete spin model.
For a finite system, this is easily implemented numerically, and we have evaluated $\widetilde{G}(r)$ for different one- and two-dimensional lattices studied here.

As expected, static mean-field theory gives a phase transition rather than a smooth crossover.
The mean-field Curie temperature $T_{\rm C}^{\rm (MF)}$ of the finite system very much depends on the size and the geometry of the underlying lattice.
If interpreted as $T_{\rm C}(L)$, the mean-field Curie temperature $T_{\rm C}^{\rm (MF)}$ yields a strong overestimation.
Here, however, our question is whether besides $T_{\rm C}^{\rm (MF)}$ there is a crossover temperature at which the spin-correlation function changes qualitatively from convex to concave.

The spin-correlation function can be obtained in two ways, either directly by computation of the thermal average $\langle \ff S_i \ff S_j \rangle$ or as the response of the local magnetic moment at site $i$ to a local magnetic field at site $j$, i.e.\ $\partial \langle \ff S_i \rangle / \partial B_j$.
In principle, both ways are equivalent because of the fluctuation-dissipation theorem $\langle \ff S_i \ff S_j \rangle - \langle \ff S_i \rangle \langle \ff S_j \rangle = T \partial \langle \ff S_i \rangle / \partial B_j$.
This is not respected by static mean-field theory, which just neglects non-local correlations.
Therefore, spin correlations can be addressed via the linear-response relation $G(\ff r)=T \partial \langle \ff S_i \rangle / \partial B_j$ only.
For an infinite translationally invariant lattice, this yields the Ornstein-Zernike form for the correlation function.
Here, for finite systems, we determine $\langle \ff S_i \rangle$ numerically by solving the static mean-field equations and compute the derivative with respect to $B_j$ numerically.
Looking at the resulting averaged correlation function $\widetilde{G}(r)$, we always find a {\em convex} trend, for any system size and dimension.
This shows that our results and the crossover temperature $T_{\rm C}(L)$ cannot be captured by a mean-field or Landau approach and therefore represent a correlation effect for which simple coordination-number arguments must be taken with care.

\section{Conclusion}

As compared to infinite bulk systems, the theoretical description of collective magnetic order is more involved for nanosized materials.
Due to the finite system size there are no clear-cut regions in parameter space where ferromagnetic order is realized.
Furthermore, the magnetic state is not stable temporally and consequently the order parameter, i.e.\ the magnetization of the nanoparticle, fluctuates with a time constant that has to be compared with the (experimental) observation time.
These facts give rise to ambiguities in the definition of the Curie temperature and imply the existence of a second temperature scale, the blocking temperature, which again cannot be defined precisely.

Our studies based on different analytical and numerical techniques have demonstrated, that a meaningful definition of the Curie temperature of a finite spin system can be given that relies on the analysis of a suitably defined average spin correlation function.
Upon lowering the temperature, the correlation function changes its curvature at $T_{\rm C}(L)$.
This definition is consistent with the usual concepts and comes closest to our expectation that the Curie point is the temperature at which the correlation length exceeds the particle size.
In particular, it accounts for the delicate interplay between the (bulk) tendency to ordering and the (surface) tendency to enhance fluctuations.
We have shown that the concept can be applied to different one-, two- and three-dimensional classical spin models and that $T_{\rm C}(L)$ can be extracted with an accuracy that even allows to determine a shift exponent.

We could further demonstrate that the blocking temperature scale is accessible with a Monte-Carlo approach by performing an incomplete statistical average.
A sharp jump is visible in the average spin-correlation function at $T_{\rm b}(L)$.
However, the blocking temperature is defined with respect to an observation time only (a finite number of Monte-Carlo sweeps) and therefore represents a relative quantity.

Concluding, the combined application of exact diagonalization, Monte-Carlo, and mean-field techniques has uncovered a strikingly complex behavior of the spin correlations in nanoparticles with qualitatively different temperatures scales.
Since simple coordination-number arguments are unable to give a quantitively correct picture of the physics, the temperature trends must be seen as effects of strong spin correlations.

\acknowledgments
Support by the DFG (SFB 668, projects B3, B4, A3, and A14) and by the Cluster of Excellence ``Nanospintronics'' Hamburg is gratefully acknowledged.

\end{document}